\documentclass[a4paper,11pt]{article}
\usepackage{pos}
\usepackage{subfig}
\usepackage{wrapfig}
\usepackage{csquotes}

\title{Vertexing detectors and vertexing performance in Run 2 in ALICE}

\manuallySeparateAuthors
\author*[a]{Mattia Faggin}
\author[]{ (on behalf of the ALICE Collaboration)}

\affiliation[a]{Università and INFN Padova,\\
  Via Marzolo 8, Padova, Italy}

\emailAdd{mattia.faggin@cern.ch}

\abstract{
The ALICE apparatus includes an Inner Tracking System for high-precision vertexing and tracking at midrapidity. Thanks to this detector, remarkable results were obtained with Run 2 data in all collision systems studied at the LHC.
In this contribution, the role of the ITS in the event, track and secondary vertex reconstruction is highlighted, and the comparison of the resolution on the track impact parameter to the collision point in data and Monte Carlo simulations is reported. Finally, some of the main physics results obtained in pp and Pb--Pb collisions relying on the excellent ALICE vertexing and tracking capabilities are presented.}

\FullConference{%
  The Eighth Annual Conference on Large Hadron Collider Physics-LHCP2020 \\
  25-30 May, 2020\\
  online}

\begin{document}
\maketitle

ALICE \cite{JInst} is a general purpose experiment composed of a central barrel at midrapidity ($|\eta|<0.9$) and a muon arm at forward rapidity ($-4<\eta<-2.5$). The track reconstruction in the central barrel is performed with the Inner Tracking System (ITS), the Time Projection Chamber (TPC) 
and the Transition Radiation Detector (TRD). 
The ITS is the innermost detector of the central barrel, made of six cylindrical layers with different technologies: the Silicon Pixel Detector (SPD), the Silicon Drift Detector (SDD) and the Silicon Strip Detector (SSD). This is a high granularity detector, designed to cope with thousands of tracks per unit of rapidity and its main purpose is the identification and separation of heavy-flavour hadron decay vertices from the primary vertex.
Furthermore, the ITS improves the tracking performance in the central barrel and it is crucial for tracking low-momentum charged particles ($p_\mathrm{T}~\lesssim~200$~MeV/$c$) that cannot be reconstructed in the TPC due to the bending caused by the magnetic field.

\begin{figure}
	\centering
	\subfloat[][Data-driven correction on impact parameter resolution for reconstructed tracks in Monte Carlo simulations.]{\includegraphics[width=0.33\textwidth]{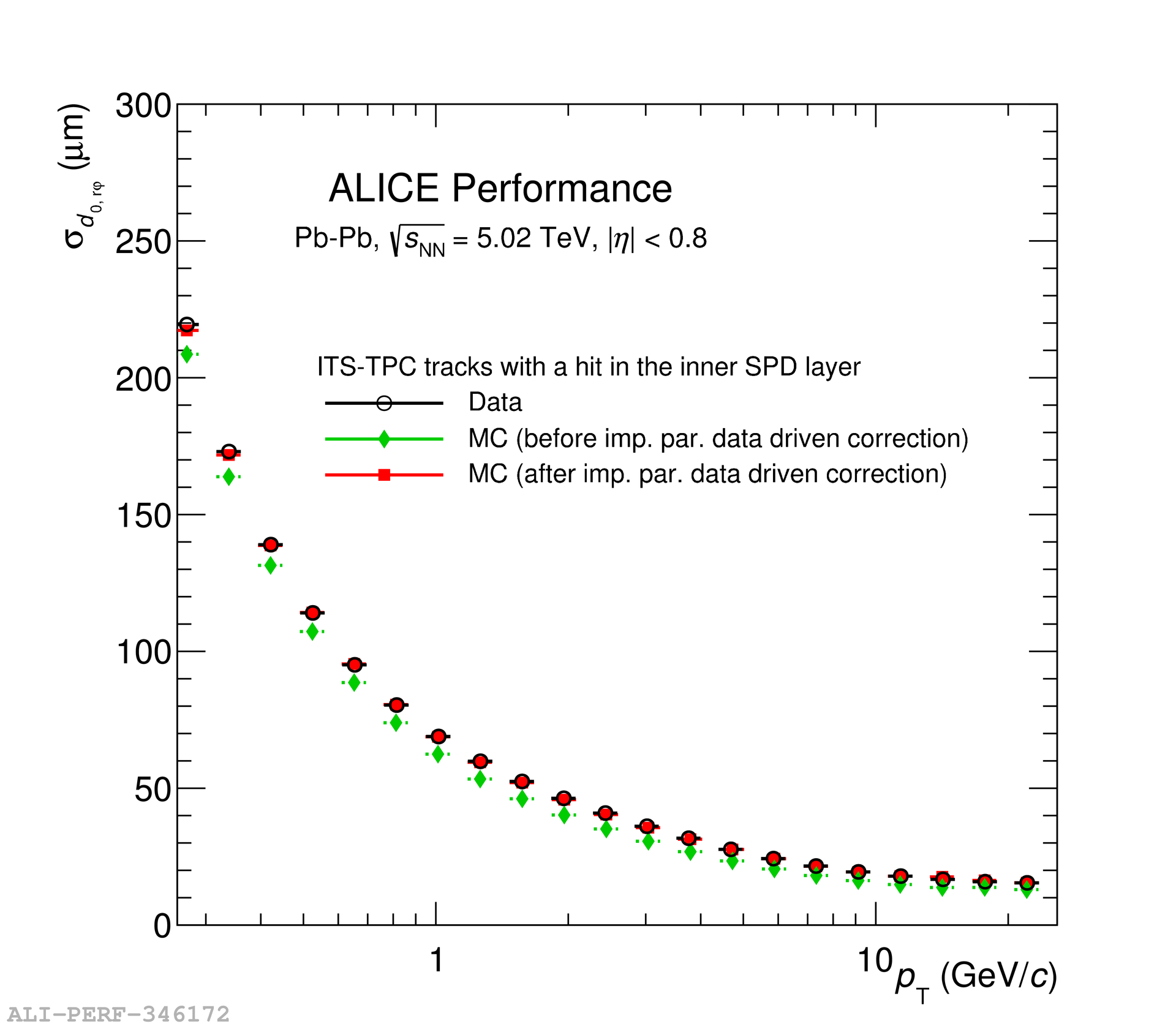}} \quad
	\subfloat[][Prompt and non-prompt D$^0$ nuclear modification factor in Pb--Pb collisions at $\sqrt{s_\mathrm{NN}}=5.02$ TeV.]{\includegraphics[width=0.28\textwidth]{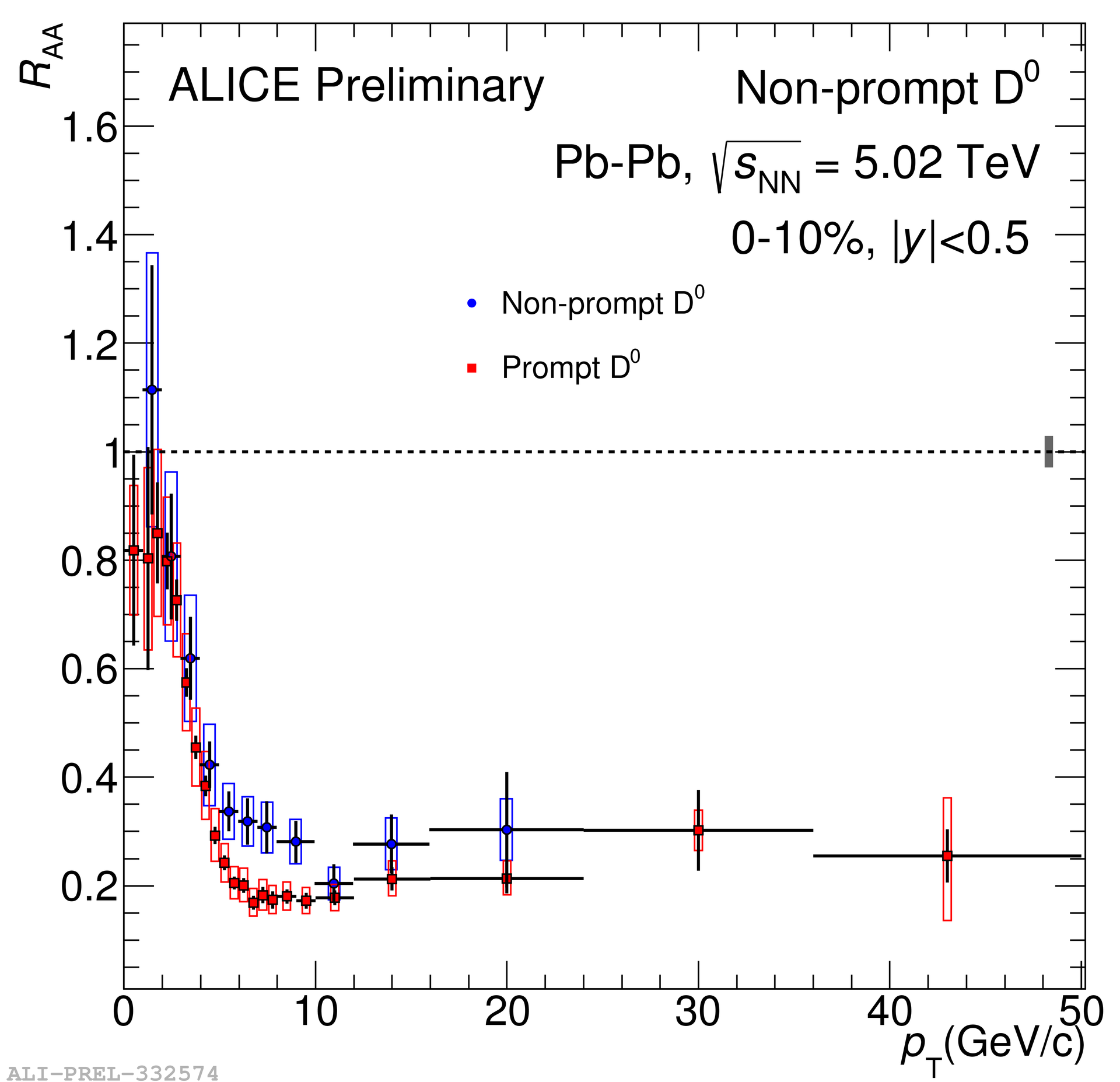}} \quad
	\subfloat[][Non-prompt J/$\psi$ fraction in pp collisions at $\sqrt{s}=13$ TeV at midrapidity.]{\includegraphics[width=0.30\textwidth]{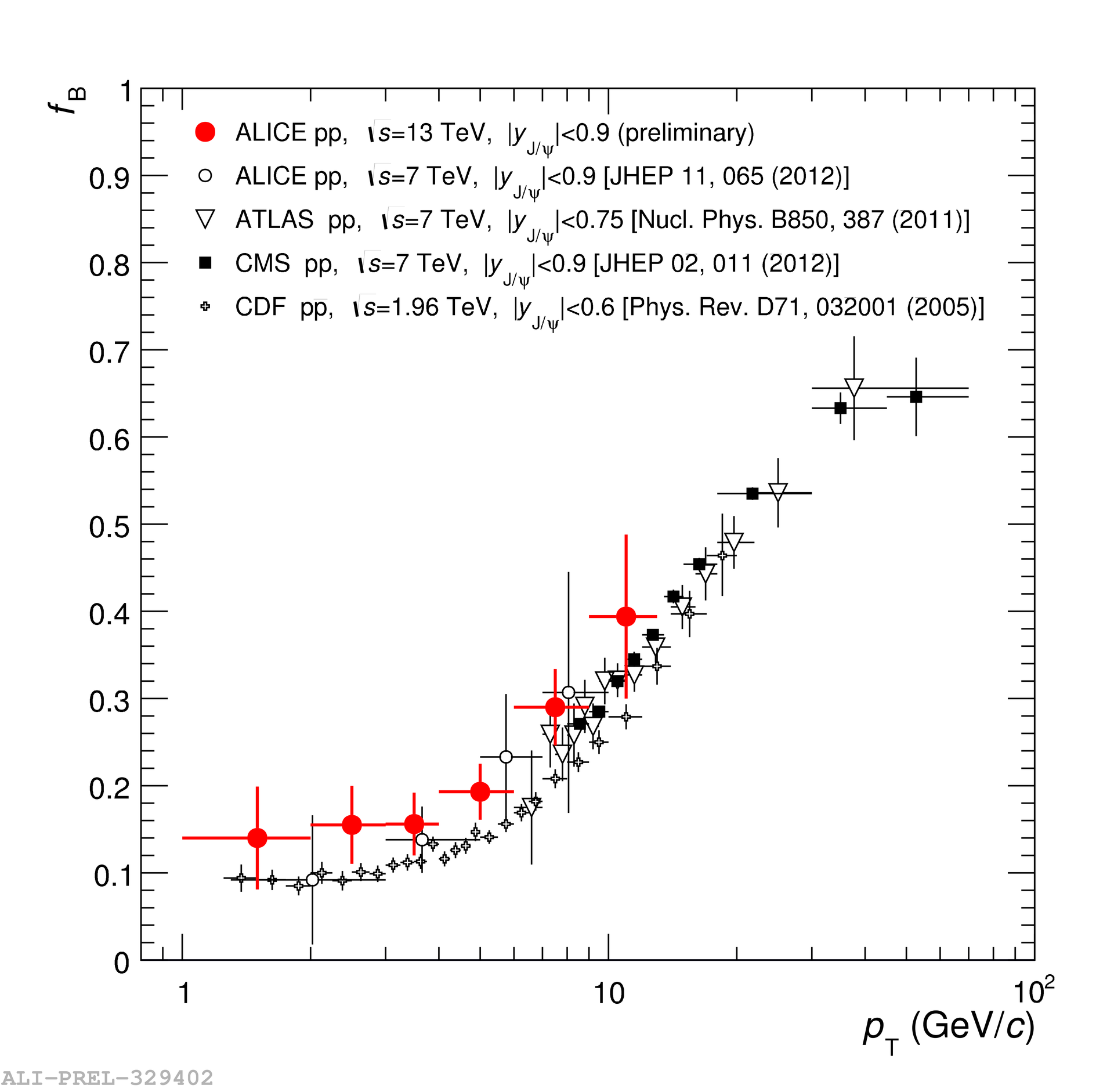}} \quad
	\caption{}
	\label{fig:nonPrompt_DmesonsRAA_nonpromptJpsi}
\end{figure}

The collisions recorded with ALICE are processed following a predefined reconstruction flow \cite{Performance}.
The procedure starts with a clusterization step, where the different detectors are treated separately and the raw data are converted into clusters defined by positions, signal amplitudes, etc., and associated errors.
Then, a preliminary reconstruction of the interaction point 
only with the SPD clusters takes place.
The track reconstruction, based on a Kalman-filter algorithm, starts from the outer radius of the TPC. 
The tracks reconstructed in the TPC are prolonged to the ITS layers, then low-momentum particles 
are searched for and tracked in the ITS. The TPC-to-ITS prolonged tracks (\enquote{global}) are propagated outwards, updated by the measurement in the TRD and then associated with clusters in the remaining detectors (TOF, HMPID, calorimeters). Finally, the global tracks undergo a further inward propagation towards the primary vertex and the track's position, direction, curvature and covariant matrix are determined. Finally, a more refined interaction-vertex finding takes place, 
then the distance to closest approach of each track to the primary vertex is calculated and the projection on the transverse plane (impact parameter) is evaluated in data and Monte Carlo simulations, as shown in Figure \ref{fig:nonPrompt_DmesonsRAA_nonpromptJpsi}(a).


Thanks to the ITS detector, ALICE is able to localize secondary vertices of heavy-flavour hadron decays by fully reconstructing the final hadronic state in the central barrel. The position is found by minimizing the distance among the daughter tracks, which are approximated as straight lines close to the primary vertex. 
The secondary vertex reconstruction is fundamental to exploit at the analysis stage several variables directly connected to the decay topology, to reduce the background from wrong track combinations and to enhance the statistical significance of the signal.
This reconstruction is performed also in Monte Carlo simulations, where the impact parameter resolution estimated with data is reproduced within 5-10~$\mu$m (Figure \ref{fig:nonPrompt_DmesonsRAA_nonpromptJpsi}(a)). Such difference, arising for instance from an imprecise description in the simulation of sensor position or orientation, needs to be corrected to avoid biases in the reconstruction and selection efficiency, and to provide unbiased templates for impact parameter -based analyses. For this reason, a correction on the track-position resolution, based on the comparison of this quantity in data and Monte Carlo simulations, is applied.

The main goal of ALICE is the study of the quark-gluon plasma (QGP) produced in heavy-ion collisions. Charm and beauty quarks are mainly produced in hard scatterings before the QGP formation, therefore they are unique and calibrated probes of the medium, since they experience its full evolution. Thanks to its vertexing capabilities, ALICE is able 
to reconstruct and identify charm-hadron decays,
despite the huge track density in the central barrel (thousands of tracks in most central Pb--Pb collisions). 
The QGP properties are investigated through several observables, such as the nuclear modification factor ($R_\mathrm{AA}$) \cite{RAA_Dmesons2015} and the elliptic flow ($v_2$) \cite{v2_beauty_electrons}, 
which are sensitive to the in-medium parton energy loss, the interplay of different heavy-quark hadronization mechanisms and the parton diffusion in the QGP. The ITS vertexing capabilities make ALICE a competitive experiment also in pp collisions, being the only LHC experiment able to measure the $\Lambda_c^+$ baryon ($c\tau=60$ $\mu$s) at midrapidity down to low $p_\mathrm{T}$ (1 GeV/$c$) \cite{LambdaC_paper_pp7TeV_pPb5.02TeV}
 , where recent ALICE measurements\footnote{These proceedings: \enquote{Recent results on hard and rare probes from ALICE}.} set significant constraints for charm hadronisation models.

\begin{figure}
	\centering
	\includegraphics[width=0.5\textwidth]{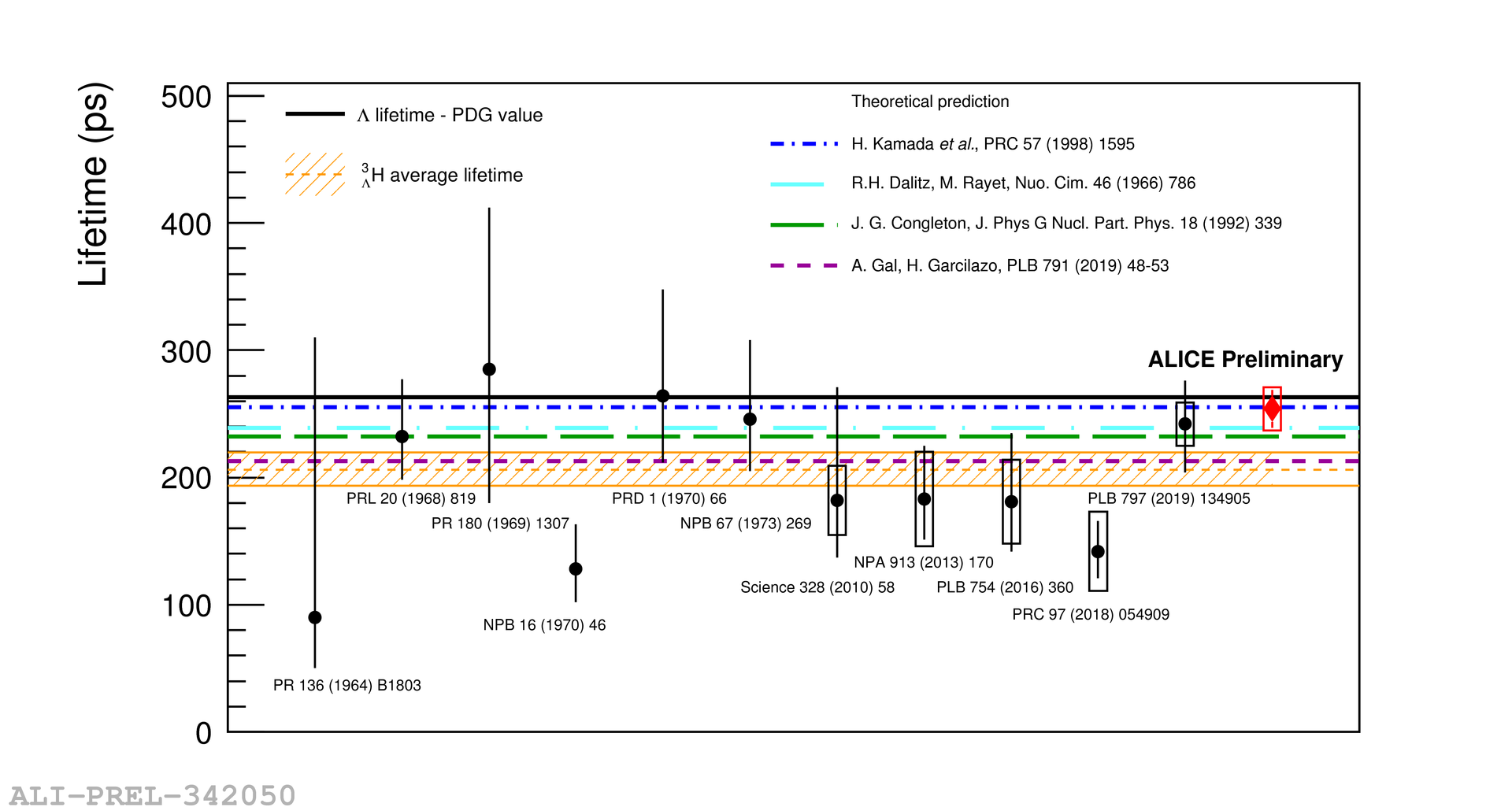}
	\caption{Compilation of $^3_\Lambda \mathrm{H}$ lifetime measurements.}
	\label{fig:hypertriton_lifetime}
\end{figure}

The impact parameter with respect to primary vertex of charm hadron decay products produced in beauty decays ($c\tau\sim 500$~$\mu$m) is larger than the one of particles produced in prompt charm hadron decays ($c\tau<$~$300$~$\mu$m). 
This kinematic property is used to disentangle electrons from beauty-hadron decays, $\mathrm{b}(\to \mathrm{c})\to e$, from charm-hadron decay electrons and from background contributions, by fitting the measured impact parameter distribution of electrons with Monte Carlo templates. This technique relies on the ITS spacial resolution (better than 35 $\mu$m in the $r\varphi$ plane \cite{JInst}) and it was used for the measurement of $\mathrm{b}(\to \mathrm{c})\to e$ elliptic flow in semi-central Pb--Pb collisions at $\sqrt{s_\mathrm{NN}}=5.02$~TeV \cite{v2_beauty_electrons}.
An additional insight about the beauty-hadron production and beauty-quark interactions with the QGP comes from the exclusive measurements of non-prompt charm hadron production, such as the production cross section of non-prompt D$^0$, D$^+$ and D$^+_s$ in pp collisions at $\sqrt{s}=13$ TeV and the $R_\mathrm{AA}$ of prompt and non-prompt D$^0$ in Pb--Pb collisions at $\sqrt{s_\mathrm{NN}}=5.02$ TeV (Figure \ref{fig:nonPrompt_DmesonsRAA_nonpromptJpsi}(b)), where an indication of $R_\mathrm{AA}^{\mathrm{D}\leftarrow \mathrm{c}}<R_\mathrm{AA}^{\mathrm{D}(\leftarrow \mathrm{c})\leftarrow \mathrm{b}}$ is observed. Such measurements are possible only by investigating different decay topologies, resolved thanks to the ALICE vertexing capabilities, exploited at their best in analyses profiting of Machine-Learning-based techniques. 
%
Further insights on the heavy-flavour production in pp collisions at midrapidity come from the measurement of prompt and non-prompt J/$\psi$ production, which are distinguished due to the sizeable displacement from the primary vertex of beauty-hadron decay point \cite{Jpsi_7TeV}. In this way, ALICE can measure the fraction of J/$\psi$ production from beauty-hadron decays in pp collisions at $\sqrt{s}=13$~TeV down to $p_\mathrm{T}=1$ GeV/$c$ (Figure \ref{fig:nonPrompt_DmesonsRAA_nonpromptJpsi}(c)) 
and the J/$\psi$ production at midrapidity, providing complementary measurements to the LHCb ones at forward rapidity. 
%
The design of the ALICE apparatus allows also for an efficient reconstruction of decay vertices of long living weakly decay particles, in particular of the $\Lambda$ baryon ($c\tau=$ 7.89 cm).
A remarkable effort in Run 2 has been dedicated to the measurement of hypertriton $^3_\Lambda \mathrm{H}$ lifetime, a fundamental brick to solve the so-called \textit{hyperon puzzle} \cite{hyperon_puzzle}. 
The lifetime is measured by reconstructing the $^3_\Lambda \mathrm{H}\to ^3$He$\pi^-$ decay vertex with the V0-finder technique described in \cite{ref_V0finder} and counting the signal in different $ct$ intervals. As visible in Figure \ref{fig:hypertriton_lifetime}, the contribution of ALICE in this field is outstanding and the precision on the lifetime measurement improved in the last years thanks to the Pb--Pb samples recorded during Run 2.

Thanks to its vertexing capabilities, ALICE obtained remarkable results in Run 2. Since the apparatus will benefit from several upgrades\footnote{These proceedings: \enquote{ALICE upgrades}.}, the ALICE Collaboration is looking forward to the next data taking campaigns in order to continue, improve and extend its physics program.


\bibliographystyle{JHEP}
\bibliography{biblio}

\end{document}